%Paper: gr-qc/9403022
%From: Tonatiuh Matos <Tonatiuh.Matos@fis.cinvestav.mx>
%Date: Thu, 10 Mar 1994 17:45:38 -0600 (CST)

\magnification\magstep1

\baselineskip 17 pt

\

\bigskip

\centerline{\bf GEODESIC MOTION IN THE 5D MAGNETIZED}
\centerline{\bf SCHWARZSCHILD-LIKE SOLUTIONS}

\vskip 2cm

\centerline{Tonatiuh Matos and Nora Bret\'on}

\centerline{Departamento de F\'{\i}sica, CINVESTAV del IPN}

\centerline{Apartado 14-740, 07000 M\'exico, D. F.}

\bigskip

{\bf ABSTRACT}

\bigskip

\hangindent = 2 cm

\hangafter = 7

Geodesics for a 5D magnetized Schwarzschild-like solution are analyzed by
reducing the problem to the motion of a test particle in an effective
potential. In absence of magnetic field comparison is established  with
Schwarzschild's geometry. Embedding diagrams are constructed in order
to visualize the geometry of the metric. The study performed here is also
valid, when the electromagnetic interactions are neglected, for the low energy
superstring theory and the Brans-Dicke theory.

\vskip 6cm


\vfill\eject

\

\bigskip

{\bf 1. Introduction}

Since many of the astrophysical objects like pulsars, quasars or black holes
are endowed with magnetic fields, it is of great interest to study solutions
of Einstein's equations which represent massive magnetic multipoles, in
particular, magnetic dipole.

However, exact solutions of Einstein's equations interpretable as massive
magnetic dipole were found only recently [1] and most of them are rather
cumbersome to be studied in analytic form.

On the other hand it is well known that the 5D solutions of Einstein's
equations in vacuum when projected into the four dimensional spacetime
correspond to solutions of Einstein's equations with an energy-momentum tensor
of a gravitational field coupled with an electromagnetic field and also with a
massless scalar field [2]. The 5D equations without sources may be reduced to
the 4D ones with sources, provided an appropriate definition is made for the
energy-momentum tensor of matter in terms of the extra part of the geometry.
In relation with this point it is not clear how to interpret any higher
dimensional theory physically. Interpretation of the fifth dimension has been
done as a massless scalar field which can be or not  associated with a fluid
density [3]. In other works it has been interpreted as a `` magnetic mass"
[4]. Also interpretation has been done as a fifth geometric property which
shows up near horizon [5]. In any case, it is worth to explore ways of
interpreting the properties of the 5D solutions in a 4D world.

 Recently, solutions were obtained using the Chiral formalism in 5D
[6]. These solutions  represent bodies endowed with a magnetic field
and whose gravitational potential possess a Schwarzschild-like behaviour. They
were derived starting with a pseudo-riemannian 5D space  which is a  gauge
invariant model under an {\bf U(1)} group. The Einstein's equations are solved
in the potential space and from these potentials the metric components are
determined.

The solution studied in this paper was derived in [7] and is a model for the
magnetic dipole of static bodies based in a 5D gauge theory that in 4D
corresponds to a massive magnetic dipole coupled to a massless scalar field.
This solution is analytically tracktable so it is worthwhile to ask about the
plausibility of such a model compared with  Schwarzschild's solution
predictions.

We note  that the analysis carried out here is valid, when the electromagnetic
field is neglected, for low energy superstring theory ($\alpha = 1$  in Ref.
[8]), for Einstein-Maxwell with scalar field ($\alpha = 0$ in Ref. [8]) and
for Brans-Dicke theory ($F = 0$ in Ref. [8]).

In the following Section 2 general features of the metric are presented. In
Section 3.1, the effective potential and the effect of variation of the
magnetic dipole are described. It is seen the effect  of the scalar field on
the spacetime near $r=2M$. In Section 3.2, in absence of magnetic field, the
precession of the perihelia of the orbit is determined for our model and
compared with the Schwarzschild's one.

In Section 3.3 the geometric character of our model is shown in diagrams of
embedding in an Euclidean three space. The fifth dimension is explored also
with the help of embedding diagrams. In fact we show in a diagram how the
extra dimension shows up near $r=2M$.

\bigskip

{\bf 2. The metric of the 5D magnetic dipole}

The metric that represents the exterior spacetime of a field with magnetic
four potential $A=A_3d\phi$ and a Schwarzschild-like gravitational potential
in the Boyer-Lindquist local coordinates $(r, \theta, \phi, t, y),$ can be
written as [7]

 $$\eqalign{ \overline g =& {1 \over {I}} \bigg\{ \sqrt{I_o g_{22}}
e^{\tau_o^, \lambda_d}  \bigg((1-{2M \over r})^{-1} dr^2  + r^2 d{\theta}^2
\bigg) \bigg({r^2-2Mr \over \Delta} \bigg)^{1 \over 3} \cr
  & + \sqrt{I_o g_{22}} r^2 \sin^2{\theta} d \phi^2  - {\big(1-{2M
\over r} \big) \over {\sqrt{I_o g_{22}}}} dt^2 \bigg\} + I^2 (A_3 d \phi +
dy)^2. \cr} \eqno(2.1)$$ where

$$ I^2={I_o \big(1-{2M \over r}\big)^{-{2 \over 3}} \over{g_{22}}}, \quad
 g_{22} = \cases{
 {} & ${a e^{q \lambda}+ b e^{-q \lambda}}$ \cr
 {} & ${b e^{iq \lambda}+ {\overline b}e^{-iq \lambda}}$ \cr
 {} & ${a \lambda + {1 \over I_o}}$ \cr}, \quad
  b+ {\overline b}={1 \over I_o}, \eqno(2.2)$$

$$\lambda={\tau_o M^2 \cos{\theta} \over {\Delta}}, \quad A_3= M^2 \tau_o (r-M)
{\sin^2{\theta} \over \Delta}, \eqno(2.3)$$

$$\Delta= (r-M)^2 - M^2 \cos^2{\theta},$$

where $\tau_o^,$ is an integration constant and  $I_o, b, {\overline b}$ and
$q$ are constants restricted to $b^2q^2$ being proportional to $I_o$. $M$ is
interpreted as the mass parameter and $\tau_o$ is the parameter associated
with the magnetic field. We observe that the scalar field $I$ appears as a
conformal factor in the metric. We interpret the term in brackets $\{ \}$
as the spacetime metric, in such a manner that the proper time $\tau$ is given
by $\{ \} = - c^2 d\tau^2.$

Like other 5D Schwarschild-like solutions, the studied metric does not have
event horizons. The curvature is singular for $r=2M$, therefore, the metric
belongs to the soliton class [9]. For $r >>M$ the electromagnetic potential
$A=A_{\mu} dx^{\mu} = A_3 d \phi$ behaves like a magnetic dipole. The metric
$(2.1)$ is asymptotically flat for $r >> M$, and flat for $M = 0$. When
$\tau_o =0$ then $A_3=0$ and $\lambda=0$ and we have the case without magnetic
field. Then the expression in brackets $\{ \}$ is like the Schwarzschild
metric. In absence of magnetic field we note from $(2.2)$ that $I^2$ is a
function that increases without bound for $ r \to 2M.$ While for $ r >> 2M$,
$I^2 \to I_o^2,$ from which we infer that the scalar potential becomes
important near $r = 2M.$

This 5D space possess three Killing vectors $ {\partial \over \partial y},
{\partial \over \partial t}, {\partial \over \partial \phi}.$ As a consequence
of the symmetries of the solution, there exist constants of motion which
enable us to reduce the problem of finding the  geodesics to the analysis of
the one-dimensional motion of a particle in an effective potential.

In order to study the geodesic motion in this space, we study the variation of
the Lagrangian

$$\eqalign{ {\cal L} = & {1 \over 2I} \bigg\{ \sqrt{I_o g_{22}} e^{\tau_o^,
\lambda_d} \bigg( (1- {2M \over r})^{-1} \big( {dr \over d\tau}\big)^2 + r^2
\big({d\theta \over d\tau}\big)^2 \bigg) \bigg({r^2-2Mr \over \Delta}
\bigg)^{1 \over 3} \cr
& +\sqrt{I_o g_{22}} r^2 \sin^2{\theta}
\big({d\phi \over d\tau}\big)^2 -{(1- {2M \over r}) \over \sqrt{I_o
g_{22}}} \big({dt \over d\tau}\big)^2 \bigg\} \cr
& + I^2A_3^2 \big({d\phi \over
d\tau}\big)^2 + 2I^2A_3 \big({dy \over d\tau}\big) \big({d\phi \over
d\tau}\big) +I^2\big({dy \over d\tau}\big)^2,    \cr} \eqno(2.4)$$

where $\tau$ is the proper time in the reference system comoving with the test
particle. We restrict to the evolution in the equatorial plane by making
$\theta = {\pi \over 2}.$  The geodesic equation is obtained from the
Lagrangian by varying ${\cal L}$ with respect to the cyclic coordinates
$\theta, t$ and $y$,

$$\eqalign{{\delta {\cal L} \over\delta \phi} \to &{ \bigg({\sqrt{I_o
g_{22}}r^2 \over{I}} + I^2 A_3^2 \bigg)} {d\phi \over d \tau}  + I^2 A_3 {dy
\over{ d \tau}}  = B, \cr
 {\delta {\cal L} \over{ \delta t}} \to & { {\big({1- {2M \over r}} \big)}
\over {I{\sqrt{I_o g_{22}}}}} {dt \over{ d \tau}} = A, \cr
 {\delta {\cal L} \over{ \delta y}} \to  & {I^2 A_3 {d \phi \over{ d \tau}} }
+ I^2 {dy \over{ d \tau}} = \hat p, \cr} $$

 $A$ and $B$ are the motion constants associated with conservation of energy
and angular momentum, respectively. We have as geodesic equation

 $$\big({dr \over d\tau} \big)^2 + e^{\big({-{\tau_o^2 \tau_o^,} \over 8(r-
1)^4} \big)} \bigg\{ \big({ {I^2(B - {\hat p}A_3)^2 \over { I_o g_{22} r^2}}
+ {c^2 \over {\sqrt{I_o g_{22}}}}} \big) \big( {1 - {2M \over r}}\big) - A^2
I^2 \bigg\}\bigg({ \Delta \over {r^2-2Mr}} \bigg)^{1 \over 3} = 0,
\eqno(2.5)$$

{}From the previous equation we have for the effective potential the expression

$$ V = \exp{\big({-{\tau_o^2 \tau_o^,} \over 8(r-1)^4} \big)} \big({ {I^2(B
- {\hat p}A_3)^2 \over { I_o g_{22} r^2}} + {c^2 \over {\sqrt{I_o g_{22}}}}}
\big) \big( {1 - {2M \over r}}\big)\bigg({\Delta \over {r^2-2Mr}} \bigg)^{1
\over 3}. \eqno(2.6)$$

\bigskip

{\bf 3. Testing the model}

\bigskip

{\bf 3.1 The effective potential for a test particle}

To deduce the qualitative features of the orbits we analyze the behaviour of
the effective potential $(2.6)$ for a test particle with angular momentum per
mass $l$ and energy $E_o$ (at infinity) in presence of our 5D massive magnetic
dipole with mass M and dipole parameter $\tau_o.$  Substituting $A_3$ from
$(2.3)$ with $\theta= {\pi \over 2}$, $\gamma = \hat p \tau_o$ and $l={B \over
M}$    we arrive to

$$V=  \exp{\big({-{\tau_o^2 \tau_o^,} \over 8(r-1)^4} \big)} \big[ \big({l
\over r} -{\gamma \over {(r-1)r}} \big)^2
+ \big(1- {2 \over r}\big)^{2 \over 3} \big] \big({1- {2 \over r} + {1 \over
r^2}} \big)^{1 \over 3}, \eqno(3.1)$$

where in order to make the plotting we have scaled ${r \over M} \to r$ and
$c=1$. For the case in absence of magnetic field we put in $(3.1)$, $\gamma
=0$ ($\tau_o =0$).

It can be seen that the last term in $(2.5)$ is proportional to
$({1 - {2M \over r}})^{-{2 \over 3}}\big({\Delta \over {r^2-2Mr}} \big)^{1
\over 3}$. In order to perform a comparison with Schwarzschild's solution, we
have taken in expression $(3.1)$ the energy $E = E_o (1-{2M \over r})^{-1}(1 -
{2M  \over r} +{M^2 \sin^2 \theta \over r^2})^{1 \over 3}$ to zeroth order. It
invalids our approximation for the region $ r \to 2 M$ because at this points
the energy diverges.

In Fig. $1$ we can visualize the principal features of typical orbits and the
effects of varying the magnetic dipole parameter $\gamma.$ The quantity
plotted is $\sqrt{1+V}.$

Fig. $1.a.$ shows, in absence of magnetic dipole ($\tau_o = 0$), the
effect on the effective potential of the variation of the angular momentum $l$
of the test particle. We take as reference for the values of the angular
momentum  of the test particle the study by Misner, Wheeler and Thorne (see Ch
25, Ref [10]). Fig. $1.a.$  can be compared with the effective potential
corresponding to Schwarzschild's solution (Fig. 25.2 in Ref [10]). It reveals
the following features: stable circular orbits are possible for the same
values of the angular momentum $l$, but the radious value of the stable orbits
are greater for the 5D model than for Schwarzschild's solution. For the 5D
model the last stable circular orbit corresponds to $l=2.03$ while for
Schwarzschild's solution this value is $l=3.464$.

Qualitatively the effect is the same as in Schwarzschild's geometry: the
minimum  (stable circular orbits) in the 5D model is ``dragged into the
hole", i. e. it is moved inwards for $l$ decreasing. However, in the 5D model
there are no a maximum (unstable circular orbits) and for the region near
$r=2M$, the effect of the scalar field  appears as a ``core" which rejects
particles which in Schwarzchild's geometry would penetrate the hole. In Fig.
$1.b.$ both Schwarzschild and 5D effective potentials are displayed for $l=4.$

In Fig. $1. c.$ it is shown the energy $E$ which we have taken to zeroth order
and the effective potential $V$ corresponding to $l=4.33.$  It occurs that the
term $A^2 I^2$ which we associate with the energy of the test particle, is not
constant and in fact it increases without bound near horizon. We can attribute
this effect to the interaction of the particle with the massless scalar field
making the particle augment its velocity in such a way that it can not be
trapped by the body. The effect of this energy would be like an impenetrable
core for a positive energy (neglecting the quantum tunnel effect). It maybe
can fit with a fluid like interpretation of the fifth dimension. In Fig.
$1.d.$ it is shown the radial velocity of the particle, $ \dot r = (E - V)^{1
\over 2},$ this velocity augments as the particle gets nearer the body and
when the particle is about $r=2M$ it suddenly increases without bound.

Turning now to the magnetic case in  $(3.1),$  Fig. $1. e.$  shows the
dependence of the effective potential $V$ with respect to the variation of
the magnetic dipole parameter $\tau_o$. In this case the value of $\tau_o$
influences in a significative way the form of the effective potential. The
effect of the magnetic dipole is to low the potential barrier. This means that
as $\tau_o$ increases, the energy for a charged particle to be captured
decreases.  Also, the effective potential is sensitive to changes of the
parameter $\gamma.$ From its definition as $\gamma={\hat p \tau_o }$, where
$\hat p$ is the fifth-momentum of the test particle, the value of $\gamma$ can
not be prescribed in advance, but rather it needs to be adjusted according to
observational data. In Fig. $1. f.$ it is compared the effective potential $V$
for a magnetic dipole with $\gamma=10$ and $\gamma=50$,  $\tau_o^2 \tau_o^,
= 1,$  with the one in absence of magnetic field.

In [11]  a solution for a massive magnetic dipole in four dimensions is
studied. Solution in [11] is singular in polar caps and therefore test
particles can never reach polar caps. For the solution analyzed here this
effect does not exist.  It can be seen from the expression for $A_3$ in
$(2.3)$ that the magnetic dipole vanishes at polar caps ($\sin{\theta} = 0$).

In the previous plots, it is manifest that the capture into the body can occur
whenever the energy of the test particle exceeds the effective potential
$V$. However if we take into account the term $E$ then no particle can be
captured by the body. It can be considered as an effect of the scalar field
surrounding the body.

\bigskip

{\bf 3.2 Precession of the perihelia of the orbit.}

In order to determine the prediction of our model  for the precession of the
perihelia of the orbit we use the equation for $r=r(\phi)$. We perform the
change of variable

$$ r={1 \over u}, \quad {dr \over d\tau} = {dr \over d\phi} {d\phi \over
d\tau} = u' {d\phi \over d\tau}. \eqno(3.2)$$

 Substituting in $\big\{ {ds \over d \tau} \big\}^2 = {-c^2}$, the geodesic
equation is

$${u'^2+e^{- \tau_o^, \lambda_d} \bigg[ u^2 (1-2Mu) -
{A^2I_og_{22} \over (B-pA_3)^2}+{c^2(1-2Mu) \sqrt{I_og_{22}} \over {I^2(B-
pA_3)^2}}\bigg]{(1- Mu)^{2 \over 3} \over {(1 - 2Mu)^{ 1 \over 3}}}}=0,
\eqno(3.3)$$
where $A$ and $B$ are the constants of motion and
$$\lambda_d = {{\tau_o^2 M^4} \over {8(r-M)^4}}, \quad A_3 = {{{M^2 \tau_o}
\over {r-M}}} $$  When $(3.3)$ is considered in the form $u'^2+V=0,$  $V$ is
the expression in square brackets. Neglecting terms in $u^3$ we have,

$$u''+ {\partial V \over{\partial u}}  = 0, \eqno(3.4)$$

with
$${\partial V \over{\partial u}}={1 \over k}- \omega^2 u +\omega_1 u^2,
\eqno(3.5)$$
with the frequencies $\omega, \omega_1$ defined by

$$\omega^2= {-{3 \gamma_1+2M \over k}+1+ \alpha \big({53M \over 15}-\gamma_1
\big) - {A^2M^2 \over 3B^2}}, $$

$$ \omega_1= {{M^2 + 3(M+\gamma_1)(M+2\gamma_1)} \over k} + 3M- \alpha
\big({182M^2 \over 45}+ {113M \gamma_1 \over 15} -3 {\gamma_1}^2 \big)
-{85A^2M^3 \over 27B^2}, \eqno(3.6)$$

and $$k={1 \over {\alpha +\gamma_1 ({A^2 \over B^2} -{3 \alpha \over{5 M}})}},
\quad \alpha={5c^2M \over{3I_o^2B^2}}, \quad \gamma_1 = {\hat p \tau_o \over
B}, $$

Now we have to solve $(3.4)$ in order to comprehend the geodesic motion of a
test particle. Solving it in a first order approximation we obtain

$$u_o={1 \over k} \big({1 \over \omega^2} + \epsilon \cos{\omega \phi} \big),
\eqno(3.7)$$ with $\epsilon$ being the excentricity of the orbit. We then
replace this result in $u^2$ in $(3.4)-(3.5)$ and solving it we obtain the
solution

$$u_1={1 \over k\omega^2} \big[1+ \epsilon \cos{({\omega_1 \over k \omega} -
\omega) \phi} \big], \eqno(3.8)$$

The expressions for the frequencies from $(3.6)$ with $\tau_o
=0$ (absence of magnetic field) reduce to

$$\omega^2= 1+ {23 \over 15} \alpha M -{A^2M^2 \over 3B^2}, \quad \omega_1 =
3M - {2 \over 45} M^2 \alpha -{85A^2M^3 \over 27 B^2}, \eqno(3.9)$$

{}From the previous expression the precession of the perihelia  can be
calculated as
$$\Delta \phi = 2 \pi \big(1-{k \omega \over{\omega_1-k \omega^2}} \big) rad.
\eqno(3.10)$$
If, for instance, we try for Mercury, we obtain a precession of $43.01'' $
(sec/century). The approximation is in good agreement with the observed data
as well as with the Schwarzschild's prediction. It is easy to show that the
constants in the final expression $(3.9)$ involve only the mass of the Sun
$M$ and the constant $k= \alpha^{-1} = (1- \epsilon^2)a$, with $a$ being the
semimajor axis and $\epsilon$ the excentricity of the eliptic orbit of
Mercury. So, at this point we haven't made any assumption about the magnitude
of the fifth dimension.

\bigskip

{\bf 3.3 Embedding}

  In order to visualize the geometry of the space around this 5D
Schwarzschild--like body in a convenient manner, we show the corresponding
embedding diagrams. For the line element $(2.1)$ in absence of magnetism,

 $$ds^2 = {1 \over I_o} \big(1- {2M \over r} \big)^{1 \over 3} \big\{
\big({{1- {2M \over r}} \over {1 -{2M \over r}+{ M^2 \sin^2 \theta \over
r^2}}} \big)^{1 \over 3} \big( {dr^2 \over {1- {2M \over r}}} + r^2
d\theta^2 \big) +r^2 \sin^2 \theta -
\big(1- {2M \over r} \big) dt^2 \big\} +$$

$$ {I_o^2 dy^2 \over {\big(1- {2M
\over r} \big)^{2 \over 3}}}. \eqno(3.11)$$

The geometry that results of making $t=const.$, $\theta={\pi \over 2}$ and
$y=const.$ is the line element

 $$ds^2 = {1 \over I_o} \big(1- {M \over r} \big)^{-{2 \over 3}} \big(1- {2M
\over r} \big)^{-{1 \over 3}}  dr^2
 + {r^2 \over I_o} \big(1- {2M \over r} \big)^{1 \over 3} d\phi^2.
\eqno(3.12)$$

We proceed to embed it in the flat geometry of an Euclidean three dimensional
manifold. We take the Euclidean three space in cylindrical coordinates $(z, r,
\phi)$ and we identify the $z$ and $\phi$ of the slice obtained from $(3.11)$
with the $z$ and the $\phi$ of the Euclidean 3-space. Then the line element in
the two cases are to be identical (see Ref [10], Ch. 23). The line element

 $$ds^2= \big[{1+ \big({dz(r) \over dr} \big)^2} \big] dr^2 + r^2 d \phi^2,
\eqno(3.13)$$

on the two-dimensional locus in the 3-geometry is identified with

 $$ds^2= f(R)dR^2 +R^2 d\phi^2, \eqno(3.14) $$

from $(3.12)$ in our body, where

$$f(R)= {{\gamma_o^6R^8} \over {(r^3-{5 \over 3}r^2)^2 (r^3 -r^2)^{2 \over
3}}}, \quad R= r \gamma_o^{-{1 \over 2}} \big({1- {2 \over r}}\big)^{1 \over
6}, \quad \gamma_o={I_o \over M^2}, \eqno(3.15)$$

here we have scaled $r \to {r \over M}$. Comparing  $(3.13)$ and $(3.14)$ it
is obtained the embedding formula

$$1+\big({dz \over dR}\big)^2= f(R), \quad dz=\sqrt{ \{f(R)-1 \}}dR.
\eqno(3.16)$$

The embedded surface  is a segment of a paraboloid of revolution. It is shown
in Fig. $2.c.$  $R$ is rescaling the radial dimension $r$. The location of the
singularity now depends on the value of the constant $\gamma_o$, associated
with the massless scalar field (see (3.15)). Making $\phi$ to vary in $(0, 2
\pi)$ we obtain  Fig. $2.a.$   Fig. $2.b.$ shows the corresponding embedding
for Schwarzschild metric. The embedding diagram in Fig.$2.a$. resembles
Schwarzschild's geometry with a ``wormhole" that could connect two distinct
asymptotically flat universes.

{}From Figs. $2.c.$ and $2.b.$ we can compare $z(R)$ from 5D with $z(r)$ for
Schwarzschild's solution. $2.c.$ is plotted for $\gamma_o = 1.3$.

We now turn to the embedding of the fifth dimension.  The exploration we
perform of the fifth dimension here is in the line of Ref [5], i. e. as a
geometric property which corresponds to an inner symmetry, rather than the
usual approach which takes its fourth-dimensional interpretation as a massless
scalar field.
 Performing a similar identification for the fifth dimension $I$ instead of
$r$ and $y$ instead of $\phi$, taking $y$ as the 5D angle variable, we have
to identify the 3-geometry in $(3.13)$ with  the slice obtained from the line
element $(3.11)$ for $t=const.$, $\phi=const.$ and $\theta = {\pi \over 2}$,

$$ds^2={{36  I^7 dI^2} \over{\gamma_o (I^3 -1)^4 [I^3+{1 \over 4}(I^3 -
1)^2]^{1 \over3}}}+I^2 dy^2 = g(I) dI^2 +I^2 dy^2,\eqno(3.17)$$

{}From $(3.13)$ and $(3.17)$ we obtain the embedding formula,

$$dz= \sqrt{\{g(I)-1\}} dI, \eqno(3.18)$$

where we have scaled to ${I \over I_o}$. A real embedding is obtained for
values of $\gamma_o$ from $10^{-6}$ and lower. Since this behaviour is
depending on values of the constant $\gamma_o = {I_o \over M^2}$ it seems that
we can not have the two embeddings simultaneously, i. e. we must choose which
dimension  we want to see, but we can not see both at the same instant when we
approach $r=2M$. This could be an artificial effect due to the form in which
we have to define $R$ for the embedding. From $(3.15)$ it can be seen that
only for $\gamma_o > 1$ (when $R > r$) we can obtain for $R$ a
real embedding. This embedding is shown in Fig. 3.

Fig. 4. is a draw of how the fifth dimension would show up near $r=2M$. This
makes sense when we think of $I$ as the fifth dimensional radious, i. e. a
dimension which is contracted and shows up near $r=2M$. From $(2.2)$, $I_o$
is the value of this radious at infinity in such a manner that when one
approaches $r=2M$, this radious increases without limit.

We also studied the redshift and null geodesics . They turn out to be the same
as  for Schwarzschild's solution as can be seen from the equation  $\big\{
{ds^2 \over d\tau^2} \big\} = 0$ from which the same Schwarzschild's equation
for geodesics is obtained.

\bigskip

{\bf 4 Conclusions}

We have analyzed the behaviour of a test particle moving in the effective
potential of a 5D Schwarzschild--like model interpreted as a massive magnetic
dipole coupled with a massless scalar field. The model behaves in agreement
with Schwarzschild's solution for regions not near $r=2M$, as shows the
prediction for the precession of the perihelia of Mercury. In the neighborhood
of $r=2M$,  it changes dramatically and fifth-dimensional effects show up.
We have discussed both interpretations, as a massless scalar field and as a
fifth--dimensional geometric property of space near the singularity $r=2M$.
The constant $\gamma_o = {I_o \over M^2}$ deserves further attention in order
to clarify its value according to real data.

The study of a solution involve a lot of interesting aspects besides the ones
analyzed here. We can point out, for instance, the study of the 4D energy
momentum tensor which corresponds to this solution; the study of the behaviour
of test particles near polar caps and of future interest is the generalization
of this solution to the stationary  case in order to model in a more realistic
way a rotating astrophysical object. It is also convenient a deeper
investigation on the relationship between this model and the string theory
for low energies.

\bigskip

 {\bf Acknowledgements}

This work is partially supported by CONACyT (M\'exico).

\vfill\eject

\

\bigskip

{\bf References}

\item{[1]}  V. S. Manko and N. R. Sibgatullin, {\it J. Math. Phys.} {\bf
34}, 170 (1993).
\item {}     Ts. I. Gutsunaev and V. S. Manko, {\it Gen. Rel. Grav.} {\bf
22}, 799 (1988).

\item{[2]}  M. J. Duff, B. F.  Nilsson and C. V. Pope {\it Phys. Rep.} {\bf
130}, 1 (1986).
\item{}     R. N. Mohapatra {\it Unification and Supersymmetry} Springer-
Verlag (1992)
\item{}     T. Matos and A. Nieto {\it Rev. Mex. Fis.} Suplement No. 2 (1993)

\item{[3]}  P.S. Wesson and J. Ponce de Le\'on, {\it J. Math. Phys.} {\bf
33}, 3883 (1992).
\item{}     P. S. Wesson, {\it Phys. Lett.} {\bf B276}, 299 (1992).

\item{[4]}  P. S. Wesson, {\it Gen. Rel. and Grav.} {\bf 22}, 707 (1990).

\item{[5]}   A. Davidson and D. A. Owen, {\it Phys. Lett.} {\bf B155},
247 (1985).

\item{[6]} T. Matos, {\it ``5D Axisymmetric Stationary Solutions as Harmonic
Maps".}  {\it J. Math. Phys.} {\bf 35} (1994).

\item{[7]} T. Matos. {\it Phys. Rev.} {\bf D48}, (1994).

\item{[8]} D. Garfinkle, G. T. Horowitz and A. Strominger {\it Phys. Rev.}
 {\bf D43}, 3140 (1991).

\item{[9]} H. Liu, {\it Gen. Rel. Grav.} {\bf 23}, 759, (1991).
\item{   } T. Dereli, {\it Phys. Lett.} {\bf B161}, 307 (1985).
\item{   } D. J. Gross and M. J. Perry, {\it Nuclear Phys.} {\bf B226}, 29
(1983).

\item{[10]}   Ch. Misner, K. S. Thorne,  J. A. Wheeler {\it Gravitation} Ed.
Freeman and Co. (1973).

\item{[11]}   D. Vokrouhlicky and V. Karas, {\it Gen. Rel. Grav.} {\bf
22}, 1033 (1990)

\vfill\eject

\

\bigskip

{\bf Figure Captions}

{\bf Fig. 1} The principal features of typical orbits and the effects on the
effective potential of varying the magnetic dipole parameter $\tau_o$ are
shown. The quantity plotted is $ \sqrt{ 1 + V}$.

a). The effective potential's variations for distinct
values of the angular momentum $l$ of the test particle, $l=4.33, 4, 3.75,
3.464, 2.5$.

b). Here are displayed both Schwarzschild's and 5D effective potentials for
$l=3.75$.

c). Energy $E = E_o(1-{2 \over r})^{-1}(1-{2M \over r} +{M^2 \sin^2{\theta}
\over r^2})^{1 \over 3}$ and the effective potential $V$ for $l=4.33$ are
plotted.

d). It is shown the variation of the radial velocity of the particle, $\dot r
= (E-V)^{1 \over 2}$. It increases without bound near $r=2M$.

e). The effect on the effective potential $V$ of the magnetic dipole parameter
variation is displayed for $\gamma = 1$ and $\tau_o^2 \tau_o^, = 10, 50, 100$.

f). The cases with magnetic dipole for $\tau_o^2 \tau_o^, =1$ and $\gamma =
10, 50$ are compared with the case in absence of magnetism.

\bigskip

{\bf Fig. 2} Embedding in an Euclidean 3-space is displayed.

a). The paraboloid of revolution in (c) is rotated and reflected in the
equatorial plane to generate the ``wormhole".

b). Here it is plotted the embedding for Schwarzschild,
$z(r) = \int{ dr \over{ \sqrt{{r \over 2} - 1}}}$.

c). Embedding for 5D metric. The points are the result of the numerical
integration $z(R)= \int{ \sqrt{f(R) - 1} dR}$. With $f(R)$ from (3.15).

\bigskip

{\bf Fig. 3} The embedding for the fifth dimension $I$ is shown. In the lower
part is shown the paraboloid of revolution which results from integrating
$z(I)= \int{\sqrt{g(I) - 1} dI}$, $g(I)$  given in (3.17). Above, the same
paraboloid has been rotated and reflected in an equatorial plane.

\bigskip

{\bf Fig. 4} This is a draw which shows how the radious of the fifth dimension
$I$ changes as a function of $r$. According to (2.2), it increases from a
small value at infinity to an infinite radious near $r=2M$.

\end